\documentclass[prb,aps,superscriptaddress,twocolumn]{revtex4-1}
\usepackage{graphicx}
\usepackage{bm}
\usepackage{amsmath}
\usepackage{amsfonts}
\usepackage{amssymb}
\usepackage{epsfig}
\usepackage{color}
\usepackage{ulem}

\begin{document}

\title{Orbital tunable 0-$\pi$ transitions in Josephson junctions with noncentrosymmetric topological superconductors}

%
\author{Yuri Fukaya}
\affiliation{Department of Applied Physics, Nagoya University, Nagoya 464-8603, Japan}
\author{Keiji Yada}
\affiliation{Department of Applied Physics, Nagoya University, Nagoya 464-8603, Japan}
\author{Yukio Tanaka}
\affiliation{Department of Applied Physics, Nagoya University, Nagoya 464-8603, Japan}
\author{Paola Gentile}
\affiliation{CNR-SPIN, c/o Universit\'a di Salerno, I-84084 Fisciano (Salerno), Italy}
\author{Mario Cuoco}
\affiliation{CNR-SPIN, c/o Universit\'a di Salerno, I-84084 Fisciano (Salerno), Italy}



\begin{abstract}
We investigate the Josephson transport properties in a Josephson junction 
consisting of a conventional $s$-wave superconductor coupled to a multi-orbital noncentrosymmetric superconductor 
marked by an orbitally driven inversion asymmetry and isotropic interorbital spin-triplet pairing.
Contrary to the canonical single band noncentrosymmetric superconductor, 
we demonstrate that the local interorbital spin-triplet pairing is tied to the occurrence of sign-changing spin-singlet pair amplitude 
on different bands with $d$-wave symmetry. 
Such multi-band $d^{\pm}$-wave state is a unique superconducting configuration 
that drives unexpected Josephson effects with 0-$\pi$ transitions displaying a high degree of electronic control. 
Remarkably, we find that the phase state of a noncentrosymmetric/$s$-wave Josephson junction can be toggled between 0 and $\pi$ 
in multiple ways through a variation of electron filling, strength of the spin-orbital coupling, 
amplitude of the inversion asymmetry interaction, and junction transparency. 
These results highlight an intrinsic orbital and electrical tunability of the Josephson response and
provide unique paths to unveil the nature of unconventional multiorbital superconductivity 
as well as inspire innovative designs of Josephson quantum devices.
\end{abstract}

\maketitle

\section{Introduction}

Breaking of inversion symmetry offers an unique possibility 
for the design of unconventional superconducting phases~\cite{sigrist91} 
in noncentrosymmetric quantum materials \cite{Bauer12,Smidman17}. 
In canonical single band noncentrosymmetric superconductors (NCSs), 
the lack of inversion symmetry naturally leads to the mixing of even (spin-singlet) 
and odd (spin-triplet) parity pairing configurations~\cite{Gorkov2001}. 
The resulting degree of parity mixing is a general consequence of the strong inversion asymmetric spin-orbit coupling 
and of the structure of the pairing interaction and can be observed in bulk materials. 

In the framework of single band NCS, a lot of attention and intense research efforts 
have been devoted to determine the relative amplitude of the opposite parity pairing components 
especially for the perspective of achieving a topological superconducting phase~\cite{Mizuno2010,Yada2011PRB,Sato2011PRB,Brydon2011PRB} 
with the spin-triplet component being dominant. 
Apart from direct spectroscopic~\cite{IHSYMTS07} or thermodynamic means to access the structure of the superconducting order parameter, 
a common and powerful approach is to design junctions that contain NCS interfaced to NCS or conventional $s$-wave superconductors (SCs). 
Several proposals have been put forward to assess the nature of the NCS 
as the formation of helical Andreev bound states (ABSs) and the corresponding anomalies 
in the conductance \cite{IHSYMTS07,Vorontsov2008}, the non-local features of the crossed Andreev reflections~\cite{Fujimoto2009}, 
the distinctive marks of the temperature dependence of the critical current~\cite{Asano2011PRB} 
and the current-voltage characteristics in NCS-NCS junctions~\cite{Borkje2006}.

The phenomenology of the Josephson response in suitably designed heterostructure with NCS 
can be quite rich due to the multi-component superconducting pairing especially when they are comparable in size. 
While the emergence of $\pi$ states is typically bound to occur 
in superconductor/ferromagnet/superconductor junctions~\cite{GolubovReview2004,Buzdinrev,Bergeret2005RMP}, 
due to the extra $\pi$ shift originating from the exchange coupling in the ferromagnetic layer, the role of spin-orbit fields 
can bring additional channels for the generation and control of $0$-$\pi$ transitions. 
Indeed, a $\pi$-Josephson effect and $0$-$\pi$ transitions can be realized 
in NCS-NCS junctions with the two NCSs 
having opposite orientation of the Rashba spin-orbit field~\cite{Liu2016} 
or by interfacing nanowires with low-dimensional electronic channels having non-trivial geometric shape at the nanoscale~\cite{Francica2020}.  
An anomalous Josephson current phase relation (CPR) can be also obtained by engineering magnetic quantum-dots 
at the NCS/$s$-wave spin-singlet superconductor (SSC) interface~\cite{Sothmann2015}. 

Interestingly, even without magnetic effects, when considering a junction 
between a conventional SSC and a NCS, 
one can achieve a transition between 0- and $\pi$/2-type of CPRs 
in the SSC/NCS junction through an anomalous $\phi$-junction behavior 
by uniquely tuning the ratio between spin-singlet and spin-triplet component~\cite{Klam2014}. 
In most of these configurations it is the balance between the spin-triplet 
and spin-singlet component that determines the overall phase coherent response of the junction.

Differently from the case of single band NCSs, 
it has been recently recognized that in materials with a strong coupling 
between spin-orbital degrees of freedom the breaking of inversion symmetry can lead to unconventional pairing 
with exotic topological properties~\cite{fukaya18}. 
Indeed, for electronic systems with atomic spin-orbit and orbital Rashba couplings, 
superconducting phases with isotropic orbital-dependent spin-triplet superconductivity can display point nodes 
that are topologically protected and manifest an extraordinary reconstruction of the excitation spectra 
both in the bulk and at the edge of the SC~\cite{fukaya18}. 
Compared with the conventional Rashba spin-orbit coupling~\cite{Rashba1960}, 
it has been realized that spin-momentum locking can be achieved by a pure orbitally driven asymmetric interaction. 
The resulting orbital Rashba effect then yields chiral orbital textures and nonstandard orbital dependent spin-textures 
through the atomic spin-orbit coupling~\cite{park11,park13,kim14,hong15,kim12}. 
Remarkably, apart from the complexity of the spin-orbital polarization pattern 
in the reciprocal space arising from the interplay of the atomic spin-orbit and orbital Rashba interactions, 
the spin vector 
of the superconducting excitations display clear 
hallmarks of the interorbital spin-triplet pairing 
with unique spin-winding around the nodal points~\cite{fukaya19}. 
The substantial nonstandard of the superconducting behavior 
for this type of multi-orbital pairing configuration poses fundamental questions 
on the nature of the transport properties in a Josephson junction based 
on such NCS and in general on the role of orbital degrees of freedom in setting out the phase state of the junction.

In this paper we demonstrate that isotropic interorbital spin-triplet pairing in NCSs 
generally leads to an intricate Josephson response within the electronic phase space 
manifesting 0-$\pi$ phase transitions when considering a junction 
that contains a conventional spin-singlet $s$-wave SC. 
This behavior is imprinted in the emergence of a unique sign-changing intraorbital spin-singlet pair amplitude 
on different bands with $d$-wave symmetry. 
Due to the anisotropic and orbital-dependent sign change of the induced intraorbital spin-singlet pair amplitude in the NCS, 
the Josephson current manifests an intrinsic tendency to undergo a transition from a 0- to a $\pi$-phase state. 
We determine 
the phase diagram associated with 
the 0 and $\pi$-states in the space spanned by the strength of the atomic spin-orbit coupling ($\lambda_{\mathrm{SO}}$) 
and the orbital Rashba interaction ($\Delta_{\mathrm{is}}$) for various electron filling factor. 
Due to the subtle orbital dependence of the induced intraorbital spin-singlet pair amplitude, 
the increase of the electron filling tends to activate more orbital channels and in turn stabilize the $\pi$-phase state 
in a large portion of the $[\Delta_{\mathrm{is}},\lambda_{\mathrm{SO}}]$ parameters space. 
The temperature dependence of the maximal Josephson current has an anomalous behavior for a junction orientation 
that is parallel to the nodal direction with a low-temperature rapid upturn 
that arises due to the presence of flat surface ABSs~\cite{Josephson1,Josephson2,Josephson3,KashiwayaTanaka2000RepProgPhys}. 
A variation of the orientation leads to a dominant second harmonic contribution in the Josephson current 
originating from the zero-energy surface ABSs. 
Due to the orbital tunability, the Josephson effect can bring unique fingerprints to unveil 
the nature of unconventional multiorbital superconductivity 
as well as inspire innovative designs of Josephson quantum devices.

The structure of the paper is as follows. 
In Sec.\ II, we introduce the model Hamiltonian and the methodology to determine the Josephson current. 
Section III is devoted to the analysis of the induced intraorbital spin-singlet pair amplitude in the bulk.
Then, we present the behavior of the CPR
in Sect. IV in terms of the spin-orbital interactions 
by varying the electron filling and discuss the origin of the sign change in the Josephson current. 
Section V is devoted to the study of the temperature dependence of the maximum Josephson current.  
Finally, the discussion and the concluding remarks are presented in Sec.\ VI.

\section{Model and methodology}

In this section, we introduce the model Hamiltonian and the methodology that has been employed to calculate 
the Josephson current for the three-orbital NCS/single band $s$-wave SC junction. 

\subsection{Model Hamiltonian}

\begin{figure}[htbp]
    \centering
    \includegraphics[width=7.8cm]{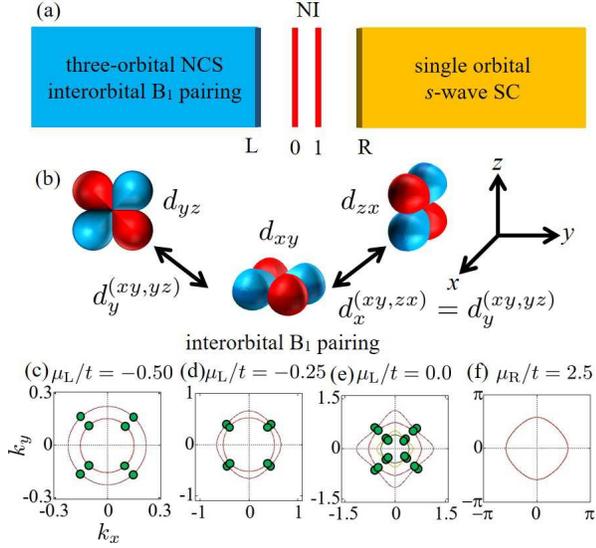}
    \caption{(a) Sketch of the noncentrosymmetric superconductor (NCS)/normal (NI)/single orbital $s$-wave superconductor (SC). 
    In the NCS, the interorbital B$_1$ pairing belongs to the C$_{4v}$ point group~\cite{fukaya18}. 
    (b) Schematic illustration of interorbital spin-triplet pairing with B$_1$ symmetry in the three-orbital NCS which is based on 
    the mixing of the $d_{xy}$ with $d_{zx}$ and $d_{yz}$-orbitals~\cite{fukaya18}. 
    (c)(d)(e)Fermi surface of NCS for $\lambda_\mathrm{SO}/t=0.10$ and $\Delta_\mathrm{is}/t=0.20$. 
    We choose three representative chemical potentials as (c)$\mu_\mathrm{L}/t=-0.50$, (d)$\mu_\mathrm{L}/t=-0.25$, and (e)$\mu_\mathrm{L}/t=0.0$.
    Filled circles denote the position of the nodes for the interorbital B$_1$ pairing. 
    (f)Fermi surface of the single orbital $s$-wave SC at the chemical potential $\mu_\mathrm{R}/t=2.5$.}
    \label{fig1}
\end{figure}

In the superconducting state we adopt a Bogoliubov-de Gennes (BdG) description. 
The left-side SC [Fig.~\ref{fig1}(a)] of the junction refers to a three-orbital NCS 
with isotropic interorbital spin-triplet pairing as schematically 
indicated in Fig.~\ref{fig1}(b). 
For this type of SC, the BdG Hamiltonian can be generally expressed in the following form
\begin{align}
\hat{H}^\mathrm{L}_\mathrm{BdG}(\bm{k})&=
\begin{pmatrix}
\hat{H}_\mathrm{L}(\bm{k}) & \hat{\Delta}_\mathrm{L} \\
\hat{\Delta}^{\dagger}_\mathrm{L} & -\hat{H}^{*}_\mathrm{L}(-\bm{k})
\end{pmatrix}.
\end{align}%
The Hamiltonian for the normal state $\hat{H}_\mathrm{L}(\bm{k})$ describes the electronic states of $d$-orbitals 
belonging to the $t_{2g}$ manifold and is given by
\begin{align}
\hat{H}_\mathrm{L}(\bm{k})&=\hat{H}_{0}(\bm{k})+\hat{H}_\mathrm{SO}+\hat{H}_\mathrm{is}(\bm{k}),
\end{align}%
with the three terms $\hat{H}_{0}(\bm{k})$, $\hat{H}_\mathrm{SO}$, 
and $\hat{H}_\mathrm{is}(\bm{k})$ ~\cite{Khalsa2013PRB,fukaya18} being associated with the orbital dependent kinetic energy, 
the atomic spin-orbit coupling, and the orbital Rashba interaction, respectively. 
The first term denotes the kinetic part,
\begin{align}
\hat{H}_{0}(\bm{k})=\hat{\varepsilon}(\bm{k})\otimes\hat{\sigma}_{0},
\end{align}%
where $\hat{\sigma}_{i=x,y,z,0}$ are the Pauli matrices and the identity matrix in the spin space.  
$\hat{\varepsilon}(\bm{k})$ corresponds to the intra-orbital kinetic energy for each $t_{2g}$-orbital,
\begin{align}
    \hat{\varepsilon}(\bm{k})&=
    \begin{pmatrix}
        \varepsilon_{yz}(\bm{k}) &0 & 0 \\
        0 & \varepsilon_{zx}(\bm{k}) & 0 \\
        0 & 0 & \varepsilon_{xy}(\bm{k})
    \end{pmatrix}, \\
    \varepsilon_{yz}(\bm{k})&=-\mu_\mathrm{L}+2t_3(1-\cos{k_x})+2t_1(1-\cos{k_y}),\\
    \varepsilon_{zx}(\bm{k})&=-\mu_\mathrm{L}+2t_1(1-\cos{k_x})+2t_3(1-\cos{k_y}),\\
    \varepsilon_{xy}(\bm{k})&=-\mu_\mathrm{L}+4t_2-2t_2\cos{k_x}-2t_2\cos{k_y}+\Delta_\mathrm{t},
\end{align}%
with $\mu_\mathrm{L}$ being the chemical potential of the NCS.
$\hat{H}_\mathrm{SO}$ expresses the canonical atomic spin-orbit coupling and is given by
\begin{align}
    \hat{H}_\mathrm{SO}&=\lambda_\mathrm{SO}[\hat{l}_{x}\otimes\hat{\sigma}_{x}+\hat{l}_{y}\otimes\hat{\sigma}_{y}+\hat{l}_{z}\otimes\hat{\sigma}_{z}],
\end{align}%
where $\lambda_\mathrm{SO}$ is the amplitude of the atomic spin-orbit interaction,
$\hat{l}_{j=x,y,z}$ are the orbital angular momentum operators in the basis $(d_{yz},d_{zx},d_{xy})$ 
projected out of the $L=2$ space. 
They are expressed as
\begin{align}
    \hat{l}_{x}&=
    \begin{pmatrix}
        0 & 0 & 0 \\
        0 & 0 & i \\
        0 & -i & 0 \\
    \end{pmatrix},\hspace{1mm}
    \hat{l}_{y}=
    \begin{pmatrix}
        0 & 0 & -i \\
        0 & 0 & 0 \\
        i & 0 & 0 \\
    \end{pmatrix},\hspace{1mm}
    \hat{l}_{z}=
    \begin{pmatrix}
        0 & i & 0 \\
        -i & 0 & 0 \\
        0 & 0 & 0 \\
    \end{pmatrix}.
\end{align}%
The third term in $\hat{H}_\mathrm{L}$ stands for the antisymmetric orbital Rashba interaction and is given by
\begin{align}
\hat{H}_\mathrm{is}(\bm{k})&=\Delta_\mathrm{is}[\hat{l}_{y}\sin{k_x}-\hat{l}_{x}\sin{k_y}]\otimes\hat{\sigma}_{0},
\end{align}%
with $\Delta_\mathrm{is}$ being the strength of the inversion symmetry breaking coupling. 
In the examined three-orbital NCS, we consider a form of interorbital local pairing that has been extensively studied in Refs.~\cite{fukaya18,fukaya19}.
There, the pair potential $\hat{\Delta}_\mathrm{L}$ can be made up by components with spin-singlet/orbital-triplet/$s$-wave and
spin-triplet/orbital-singlet/$s$-wave pairing symmetry.
Thus, the pair potential $\hat{\Delta}_\mathrm{L}$ is described by the $t_{2g}$-orbital characters $\alpha,\beta=yz,zx,xy$
for each interorbital pairing symmetry,
\begin{align}
    \hat{\Delta}^{(\alpha,\beta)}_\mathrm{L}=i\hat{\sigma}_{y}\psi^{(\alpha,\beta)}
    +i[\bm{d}^{(\alpha,\beta)}\cdot\hat{\bm{\sigma}}]\hat{\sigma}_{y},
\end{align}%
where $\psi^{(\alpha,\beta)}$ is the spin-singlet/orbital-triplet pair potential
and $\bm{d}^{(\alpha,\beta)}$ are the \textbf{d}-vectors, 
\begin{align*}
    \bm{d}^{(xy, yz)}&=\left(d^{(xy, yz)}_{x}, d^{(xy, yz)}_{y}, d^{(xy, yz)}_{z}\right), \\
    \bm{d}^{(xy, zx)}&=\left(d^{(xy, zx)}_{x}, d^{(xy, zx)}_{y}, d^{(xy, zx)}_{z}\right), \\
    \bm{d}^{(yz, zx)}&=\left(d^{(yz, zx)}_{x}, d^{(yz, zx)}_{y}, d^{(yz, zx)}_{z}\right).
\end{align*}%
The spin-triplet/orbital-singlet state for each interorbital isotropic pairing is described by the following \textbf{d}-vectors, 
\begin{align*}
    \hat{\Delta}^{(\alpha,\beta)}_\mathrm{L}&=
    \begin{pmatrix}
        \Delta^{(\alpha,\beta)}_{\uparrow\uparrow} & \Delta^{(\alpha,\beta)}_{\uparrow\downarrow}\\
        \Delta^{(\alpha,\beta)}_{\downarrow\uparrow} & \Delta^{(\alpha,\beta)}_{\downarrow\downarrow}
    \end{pmatrix}\\
    &=
    \begin{pmatrix}
        -d^{(\alpha,\beta)}_x+i d^{(\alpha,\beta)}_y & d^{(\alpha,\beta)}_z \\
        d^{(\alpha,\beta)}_z & d^{(\alpha,\beta)}_x+id^{(\alpha,\beta)}_y
    \end{pmatrix}.
\end{align*}%
In this study, we consider an interorbital pairing state belonging to the B$_1$ representation of the C$_{4v}$ point group 
[Fig.~\ref{fig1}(b)] that is the most favorable energetically
among all the allowed interorbital pairings~\cite{fukaya18}.  
This pairing state is described by a pure spin-triplet configuration and exhibits nodal points along the diagonal direction [Figs.~\ref{fig1}(c)-\ref{fig1}(e)]
which are topologically protected by the chiral symmetry of the BdG Hamiltonian~\cite{fukaya18}.
The \textbf{d}-vector of the interorbital B$_1$ pairing state is given by
\begin{align}
d^{(xy, zx)}_{x}&=d^{(xy, yz)}_{y},\notag \\
\Delta^{\uparrow \uparrow}_{xy,yz}&=\Delta^{\downarrow \downarrow}_{xy,yz}=id^{(xy, yz)}_{y},\notag \\
\Delta^{\uparrow \uparrow}_{xy,zx}&=-\Delta^{\downarrow \downarrow}_{xy,zx}=-d^{(xy, zx)}_{x}.
\end{align}%
We point out that a different \textbf{d}-vector orientation is associated with the interorbital pairing 
when mixing the $(d_{xy},d_{zx})$ or $(d_{xy},d_{yz})$ orbitals.

On the other hand, for the description of the right-side SC in the junction we consider a canonical single orbital $s$-wave state, 
\begin{align}
\hat{H}^\mathrm{R}_\mathrm{BdG}(\bm{k})&=
\begin{pmatrix}
\hat{H}_\mathrm{R}(\bm{k}) & \hat{\Delta}_\mathrm{R} \\
\hat{\Delta}^{\dagger}_\mathrm{R} & -\hat{H}^{*}_\mathrm{R}(-\bm{k})
\end{pmatrix}.
\end{align}%
Here, $\hat{H}_\mathrm{R}(\bm{k})$ denotes the Hamiltonian in the normal state for the single orbital model,
\begin{align}
\hat{H}_\mathrm{R}(\bm{k})&=\xi_\mathrm{R}(\bm{k})\otimes\hat{\sigma}_{0},\\
\xi_\mathrm{R}(\bm{k})&=-\mu_\mathrm{R}+4t_4-2t_4\cos{k_x}-2t_4\cos{k_y},
\end{align}%
with $\mu_\mathrm{R}$ being the chemical potential of the single orbital $s$-wave SC. 
The pair potential $\hat{\Delta}_\mathrm{R}$ is given by
\begin{align}
\hat{\Delta}_\mathrm{R}&=i\hat{\sigma}_{y}\psi_\mathrm{R},
\end{align}%
with the spin-singlet/$s$-wave pair potential $\psi_\mathrm{R}$.

In the normal layers between the two SCs, 
we consider the following single orbital model Hamiltonian, 
\begin{align}
\hat{H}_\mathrm{NI}(\bm{k})&=\xi_\mathrm{NI}(\bm{k})\otimes\hat{\sigma}_{0},\\
\xi_\mathrm{NI}(\bm{k})&=-\mu_\mathrm{NI}+4t_5-2t_5\cos{k_x}-2t_5\cos{k_y},
\end{align}%
with $\mu_\mathrm{NI}$ being the chemical potential setting the electron density at the normal insulating layer.

In our calculation, we set the parameters as $t_2=t_1=t_4=t_5=t$, $t_3=0.10t$, $\Delta_\mathrm{t}=-0.50t$, 
$\mu_\mathrm{R}=2.5t$, and $\mu_\mathrm{NI}=-0.50t$.
In addition, we fix the critical temperature of the two SCs
as $T_\mathrm{cL}/t=1.0\times 10^{-5}$ and $T_\mathrm{cR}/t=10 T_\mathrm{cL}$.
Then, we assume that the gap amplitude of the SCs $\Delta_\mathrm{L}(T)$ and $\Delta_\mathrm{R}(T)$ 
has a BCS-like temperature dependence $T$,
\begin{align} 
    \Delta_X(T)&=\Delta_X(0)\tanh\left[1.74\sqrt{\frac{T_{\mathrm{c}X}-T}{T}}\right],\notag\\
    \Delta_X(0)&=\frac{3.53}{2}T_{\mathrm{c}X},
\end{align}%
where $X=\mathrm{L,R}$ denotes the index for the left and right-side SCs
within the junction, respectively.

\subsection{Recursive Green's function approach}

In order to compute the Josephson current, we employ the recursive Green's function method~\cite{LDOSUmerski}.
As shown in Fig.~\ref{fig1}(a), we consider the two semi-infinite SCs 
and two normal layers sandwiched between the SCs as studied in Ref.~\cite{KawaiPRB}. 
Firstly, we calculate the semi-infinite surface Green's functions 
for the left and right-side SCs
$G_\mathrm{L}(k_{\parallel},i\varepsilon_{n})$ and $G_\mathrm{R}(k_{\parallel},i\varepsilon_{n},\phi)$ 
with  $i\varepsilon_{n}=i(2n+1)\pi k_\mathrm{B}T$ being the fermionic Matsubara frequency, 
$\phi$ the phase difference between two SCs,
and $k_{\parallel}$ the momentum that is parallel to the interface. 
When we include the normal layers at the boundary of a SC, 
these surface Green's functions, i.e.\ $G_\mathrm{L0}(k_{\parallel},i\varepsilon_{n})$ 
and $G_\mathrm{R1}(k_{\parallel},i\varepsilon_{n},\phi)$, are given by
\begin{align} 
    &G_\mathrm{L0}(k_{\parallel},i\varepsilon_{n})
    =\left[i\varepsilon_n-\hat{u}_\mathrm{NI}-\hat{t}^{\dagger}_\mathrm{L,NI}G_\mathrm{L}(k_{\parallel},i\varepsilon_{n})\hat{t}_\mathrm{L,NI}\right]^{-1},\\
    &G_\mathrm{R1}(k_{\parallel},i\varepsilon_{n},\phi)\notag\\
    &=\left[i\varepsilon_n-\hat{u}_\mathrm{NI}-\hat{t}_\mathrm{R,NI}G_\mathrm{R}(k_{\parallel},i\varepsilon_{n},\phi)\hat{t}^{\dagger}_\mathrm{R,NI}\right]^{-1},
\end{align}%
with $\hat{u}_\mathrm{NI}$ setting the on-site electron density of the normal layer.
Here, $\hat{t}_\mathrm{L,NI}$ ($\hat{t}_\mathrm{R,NI}$) means the tunnel Hamiltonian 
between left (right)-side SC and the normal insulating layer, 
\begin{align}
    \hat{t}_\mathrm{X,NI}(k_\parallel)&=
    \begin{pmatrix}
        \tilde{t}_\mathrm{X,NI}(k_\parallel) & 0 \\
        0 & -\tilde{t}^{*}_\mathrm{X,NI}(-k_\parallel)
    \end{pmatrix}.
\end{align}%
In the (100) direction, these are described by
\begin{align} 
    \tilde{t}_\mathrm{L,NI}(k_{\parallel})=t_\mathrm{int}
    \begin{pmatrix}
    -t & 0 \\
    -t & 0\\
    -t & 0\\
    0 & -t \\
    0 & -t \\
    0& -t
    \end{pmatrix},
\end{align}%
\begin{align} 
    \tilde{t}_\mathrm{R,NI}(k_{\parallel})=t_\mathrm{int}
    \begin{pmatrix}
    -t & 0 \\
    0 & -t
    \end{pmatrix},
\end{align}%
and in the (110) direction,
\begin{align} 
    \tilde{t}_\mathrm{L,NI}(k_\parallel)=t_\mathrm{int}
    \begin{pmatrix}
    t(k_\parallel) & 0 \\
    t(k_\parallel) & 0 \\
    t(k_\parallel) & 0 \\
    0 & t(k_\parallel)\\
    0 & t(k_\parallel)\\
    0 & t(k_\parallel)\\
    \end{pmatrix},
\end{align}%
\begin{align} 
    \tilde{t}_\mathrm{R,NI}(k_\parallel)=t_\mathrm{int}
    \begin{pmatrix}
    t(k_\parallel) & 0 \\
    0 & t(k_\parallel)
    \end{pmatrix},
\end{align}%
with $t(k_\parallel)=-2t\cos{k_\parallel}$ and $t_\mathrm{int}$ setting the degree of the junction's transparency.
Next, when connecting two SCs with a normal layer as shown in Fig~\ref{fig1}(a), 
one can calculate the local Green's functions $G_\mathrm{00}(k_{\parallel},i\varepsilon_{n},\phi)$ 
and $G_\mathrm{11}(k_{\parallel},i\varepsilon_{n},\phi)$,
\begin{align} 
    G_\mathrm{00}(k_{\parallel},i\varepsilon_{n},\phi)
    &=\left[G^{-1}_\mathrm{L0}(k_{\parallel},i\varepsilon_{n})
    -\hat{t}_\mathrm{NI}G_\mathrm{R1}(k_{\parallel},i\varepsilon_{n},\phi)\hat{t}^{\dagger}_\mathrm{NI}\right]^{-1},\\
    G_\mathrm{11}(k_{\parallel},i\varepsilon_{n},\phi)
    &=\left[G^{-1}_\mathrm{R1}(k_{\parallel},i\varepsilon_{n},\phi)
    -\hat{t}^{\dagger}_\mathrm{NI}G_\mathrm{L0}(k_{\parallel},i\varepsilon_{n})\hat{t}_\mathrm{NI}\right]^{-1},
\end{align}%
and the non-local Green's functions $G_\mathrm{01}(k_{\parallel},i\varepsilon_{n},\phi)$
and $G_\mathrm{10}(k_{\parallel},i\varepsilon_{n},\phi)$,
\begin{align} 
    G_\mathrm{01}(k_{\parallel},i\varepsilon_{n},\phi)
    &=G_\mathrm{L0}(k_{\parallel},i\varepsilon_{n},\phi)
    \hat{t}_\mathrm{NI}(k_{\parallel})G_\mathrm{11}(k_{\parallel},i\varepsilon_{n},\phi),\\
    G_\mathrm{10}(k_{\parallel},i\varepsilon_{n},\phi)
    &=G_\mathrm{R1}(k_{\parallel},i\varepsilon_{n},\phi)
    \hat{t}^{\dagger}_\mathrm{NI}(k_{\parallel})G_\mathrm{00}(k_{\parallel},i\varepsilon_{n},\phi),
\end{align}%
with the $\hat{t}_\mathrm{NI}(k_{\parallel})$ being the nearest-neighbor hopping term in the normal layer.
Concerning the current operator, one can calculate the Josephson current $I_\mathrm{c}(\phi)$ at a given phase difference $\phi$ 
between the left and right side of the junction by evaluating the following expression,  
\begin{align} 
    &I_\mathrm{c}(\phi)
    =\frac{ie}{\hbar}\int^{\pi}_{-\pi}\mathrm{Tr}'k_\mathrm{B}T\notag\\
    &\times\sum_{i\varepsilon_n}\left[\hat{t}_\mathrm{NI}(k_{\parallel})G_{01}(k_{\parallel},i\varepsilon_n,\phi)
    -\hat{t}^{\dagger}_\mathrm{NI}(k_{\parallel})G_{10}(k_{\parallel},i\varepsilon_n,\phi)\right]dk_{\parallel}.
\end{align}%
Here, $\mathrm{Tr}'$ means the trace over the electronic degrees of freedom. 
In this study, we focus on three representative types of spin-resolved Fermi surfaces for
the NCS at $\mu_\mathrm{L}/t=-0.50$ [Fig.~\ref{fig1}(c)], 
$\mu_\mathrm{L}/t=-0.25$ [Fig.~\ref{fig1}(d)], and $\mu_\mathrm{L}/t=0.0$ [Fig.~\ref{fig1}(e)], 
and we fix the chemical potential of the single orbital $s$-wave SC
at $\mu_\mathrm{R}/t=2.5$ [Fig.~\ref{fig1}(f)]. 
In the NCS, we consider the spin-split Fermi surfaces with both nonzero spin-orbit coupling $\lambda_\mathrm{SO}$ 
and the orbital Rashba interaction $\Delta_\mathrm{is}$.
At $\mu_\mathrm{L}/t=-0.50$ [Fig.~\ref{fig1}(c)] and $\mu_\mathrm{L}/t=-0.25$ [Fig.~\ref{fig1}(d)], 
there are two Fermi surfaces and the $d_{xy}$ is the dominant orbital component at the Fermi level. 
On the other hand, for $\mu_\mathrm{L}/t=0.0$ [Fig.~\ref{fig1}(e)], 
the number of Fermi surfaces is four and these Fermi surfaces typically include all $t_{2g}$-orbitals. 
We can thus evaluate the influence of the orbital character and the number of Fermi surfaces
by calculating the Josephson current for each selected $\mu_\mathrm{L}$.

\section{Induced intraorbital spin-singlet pair amplitude}

We start by analyzing the induced intraorbital spin-singlet pair amplitude 
for the three representative types of spin-split Fermi surfaces 
as shown in Figs.~\ref{fig1}(c)-\ref{fig1}(e)
at $\mu_\mathrm{L}/t=-0.50$, $\mu_\mathrm{L}/t=-0.25$, $\mu_\mathrm{L}/t=0.0$ 
and we will consider its profile both in the bulk and in the following section at the junction's interface. 

In the Josephson junction upon examination, the even-frequency spin-singlet pairing components 
in both left and right-side SCs can interfere
and contribute to the first harmonic term of the overall Josephson current. 
For this reason, it is useful to investigate the spin-singlet components of the pair amplitude on the Fermi surfaces
in the NCS both in the inner side at a given $k$ in the reciprocal space 
or along the edge of the junction's interface for the conserved component of the momentum. 
Hereafter, we indicate as $F^{(\alpha,\beta)}_{\uparrow\downarrow-\downarrow\uparrow}(k)$ the spin-singlet pair amplitude associated with 
the electron pairing in the orbitals $(\alpha,\beta)$ at a given value of the momentum $k$.

\begin{figure}[htbp]
    \centering
    \includegraphics[width=7.8cm]{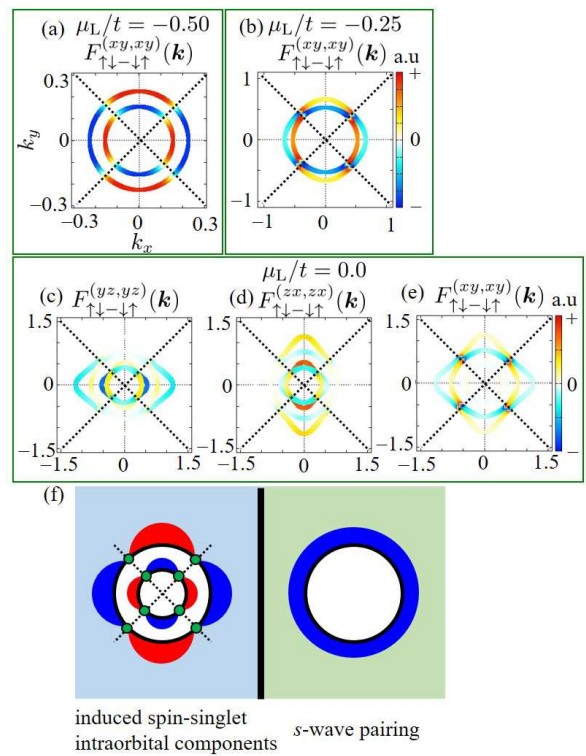}
    \caption{Even-frequency spin-singlet intraorbital pairing amplitude on the Fermi surfaces with $d_{xy}$ character evaluated 
    in the bulk of the NCS at (a)$\mu_\mathrm{L}/t=-0.50$ and (b)$\mu_\mathrm{L}/t=-0.25$. 
    Spin-singlet intraorbital pairing amplitude with (c)$d_{yz}$, (d)$d_{zx}$, and $d_{xy}$ orbital character 
    in the NCS bulk at $\mu_\mathrm{L}/t=0.0$. 
    All even-frequency spin-singlet intraorbital components have the $d^{\pm}_{x^{2}-y^{2}}$-wave structure with sign change 
    when comparing with the inner and outer Fermi surfaces. 
    (f)Schematic illustration of the Josephson junction. Black dotted line denotes the direction along 
    which nodal points occur while green circles stand for the position of the node. 
    We set the parameters as $\lambda_\mathrm{SO}/t=0.10$, $\Delta_\mathrm{is}/t=0.20$ 
    for the spin-orbit and orbital Rashba couplings and the temperature is $T=0.10T_\mathrm{cL}$.}
    \label{fig2_v2}
\end{figure}%
Regarding the bulk NCS, we find that at the Fermi surface, 
for the two representative values of the chemical potential $\mu_\mathrm{L}/t=-0.50$ and $\mu_\mathrm{L}/t=-0.25$, 
the intraorbital spin-singlet component associated with the $d_{xy}$ configuration 
has a sign-changing $d_{x^{2}-y^{2}}$-wave structure with nodal points along the diagonal direction
for each Fermi surface as shown in Figs.~\ref{fig2_v2}(a) and \ref{fig2_v2}(b). 
In particular, we point out that the sign of the pair amplitude on the inner Fermi surface 
is opposite as compared with that on the outer Fermi surface. 
Thus, the intraorbital spin-singlet pair amplitude realizes a $d^{\pm}_{x^{2}-y^{2}}$-wave pairing configuration 
with a band dependent sign of the pair amplitude that resembles the isotropic $s_{\pm}$-wave proposed in the framework 
of the iron based SCs~\cite{Mazin2008,Kuroki2008,Bang2017}.
Likewise, at the Fermi level $\mu_\mathrm{L}/t=0.0$, the intraorbital spin-singlet component has
also $d^{\pm}_{x^{2}-y^{2}}$-wave structure with nodal points along the diagonal direction as explicitly demonstrated in Figs.~\ref{fig2_v2}(c)-\ref{fig2_v2}(e). 
However, due to the contribution of the $d_{zx}$ and $d_{yz}$-bands, 
the momentum distribution of the pair amplitude is more anisotropic than the $d_{xy}$ case 
when considering the corresponding intraorbital configurations [Figs.~\ref{fig2_v2}(d) and \ref{fig2_v2}(e)]. 
We note that also for this $d^{\pm}_{x^{2}-y^{2}}$-wave state,
the intraorbital spin-singlet component pair amplitude 
has opposite signs on the inner and outer Fermi surface [Figs.~\ref{fig2_v2}(c)-\ref{fig2_v2}(e)].
Thus, the induced $d^{\pm}_{x^{2}-y^{2}}$-wave pairing, as schematically shown in Fig.~\ref{fig2_v2}(f), 
emerges as a relevant element to interpret and evaluate the Josephson effect especially 
when considering the junction with the NCS interfaced to a $s$-wave spin-singlet SC. 
Indeed, even if the interorbital spin-triplet pairing symmetry is dominant in the NCS, 
we expect that the induced intraorbital spin-singlet $d^{\pm}_{x^{2}-y^{2}}$-wave configuration will play a key role 
in setting the Josephson current and would naturally lead to a sign frustration in the Josephson current due to the sign effects at the Fermi surface. 
Moreover, due to the significant orbital dependence and the momentum anisotropy 
we also expect that 0-$\pi$ transitions can be sensitive to the junction transparency.

\section{Current phase relation: phase diagram, role of interface orientation, transparency and temperature}

In this section, we present the CPR for the interorbital B$_1$ state NCS/NI/single orbital $s$-wave SC junction (NCS/NI/SSC). 
The CPR can be generally expanded in Fourier series in terms of all the harmonics with respect 
to the applied phase difference $\phi$ as follows,
\begin{align}
I_\mathrm{c}(\phi)&=\sum_{n}[I_{n}\sin({n\phi})+J_{n}\cos({n\phi})].
\end{align}%
Since for the examined junction both SCs have the time-reversal symmetry, 
the cosine term $J_{n}$ equals to zero~\cite{AsanoPRB2003}.
 
\begin{figure}[htbp]
    \centering
    \includegraphics[width=7.8cm]{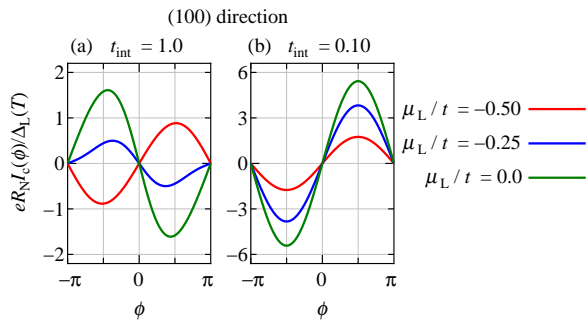}
    \caption{Current phase relation (CPR) for the NCS/NI/SSC junction
    with the interface perpendicular to the (100) direction assuming that
    $\mu_\mathrm{L}/t=-0.50$ (red line), $\mu_\mathrm{L}/t=-0.25$ (blue line), 
    and $\mu_\mathrm{L}/t=0.0$ (green line). 
    The amplitude of the spin-orbital and orbital Rashba interactions
    corresponds to $\lambda_\mathrm{SO}/t=0.10$ and $\Delta_\mathrm{is}/t=0.20$. 
    The temperature is set at $T=0.10T_\mathrm{cL}$. 
    The results correspond to two different regimes of junction's transparency: high transparency with $t_\mathrm{int}=1.0$ 
    in (a), and low transparency for $t_\mathrm{int}=0.10$ in (b). 
    We find that in the regime of high transparency there is a 0-$\pi$ transition which is obtained 
    by varying the electron filling from low to high density. 
    For the low transparent regime at the interface (i.e.\ $t_\mathrm{int}=0.10$) there is no phase change. 
    This indirectly indicates that by modifying the transparency one can drive a 0-$\pi$ transition.}
    \label{fig3_v2}
\end{figure}

    
Let us first discuss the outcome of the CPR for the (100) junction orientation.
In Fig.~\ref{fig3_v2}(a), we report the CPR assuming that the charge transfer electronic processes 
at the interface set out a regime of high transparency with the hopping amplitude $t_\mathrm{int}=1.0$. 
Hence, in order to assess the role of the orbital degree of freedom we investigate three representative chemical potentials 
for the NCS, i.e.\ $\mu_\mathrm{L}/t=-0.50$ (red line), $\mu_\mathrm{L}/t=-0.25$ (blue line), and $\mu_\mathrm{L}/t=0.0$ (green line)
in Fig.~\ref{fig3_v2}(a). 
Here, when the Fermi surface is dominated only by the $d_{xy}-$orbital,
we find that the CPR has a conventional sinusoidal $0$-junction behavior at $\mu_\mathrm{L}/t=-0.50$ (red line)
as shown in Fig.~\ref{fig3_v2}(a). 
However, with the increase of the electron filling via $\mu_\mathrm{L}$, the Josephson current relation changes to a $\pi$-phase profile 
with a sign change [Fig.~\ref{fig3_v2}(a)]. 
\begin{figure}[htbp]
    \centering
    \includegraphics[width=7cm]{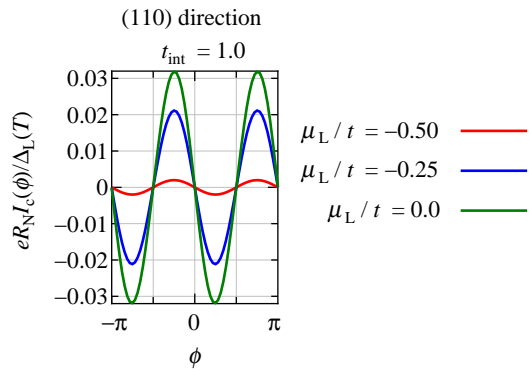}
    \caption{CPR for the NCS/NI/SSC junction in the (110) direction 
    for $\mu_\mathrm{L}/t=-0.50$ (red line), $\mu_\mathrm{L}/t=-0.25$ (blue line), 
    and $\mu_\mathrm{L}/t=0.0$ (green line). 
    We set the amplitude of the spin and orbital electronic parameters as $\lambda_\mathrm{SO}/t=0.10$, 
    $\Delta_\mathrm{is}/t=0.20$, $t_\mathrm{int}=1.0$, and $T=0.10T_\mathrm{cL}$. 
    In the (110) the second harmonic contribution dominates the Josephson current behavior.}
    \label{fig4_v2}
\end{figure}%
This trend indicates that a 0-junction can be turned into a $\pi$-junction 
by suitably tuning the band occupation through the chemical potential $\mu_\mathrm{L}$.
On the other hand, for the case of low transparency ($t_\mathrm{int}=0.10$), 
we find that the Josephson current is always conventional and no sign change is observed [Fig.~\ref{fig3_v2}(b)].


A change in the junction orientation leads to a dramatic impact on the Josephson response. 
Indeed, if we select a junction interface with (110) direction the presence of nodal points both 
in the dominant isotropic interorbital spin-triplet pairing component 
and in the induced spin-singlet $d^{\pm}_{x^{2}-y^{2}}$-wave pairing offers the opportunity 
to explore a highly nontrivial case of unconventional superconductivity. 
As for the (100) orientation, for the first harmonic term the even-frequency/spin-singlet intraorbital component in the NCS
can be coupled to the even-frequency/spin-singlet pairing in the $s$-wave SC. 
However for this case, first harmonic term $I_1$ vanishes 
since the B$_1$ pairing in the NCS is odd under the mirror symmetry along the diagonal direction,
while SSC is even.
It is the same as the case of the single band $d$-wave based superconducting junctions~\cite{Josephson1,Josephson2,Josephson3,KashiwayaTanaka2000RepProgPhys}.
Moreover, the Josephson current is substantially independent of the amplitude of the chemical potential $\mu_\mathrm{L}$ 
as demonstrated in Fig.~\ref{fig4_v2}.


\begin{figure}[htbp]
\centering
\includegraphics[width=7.8cm]{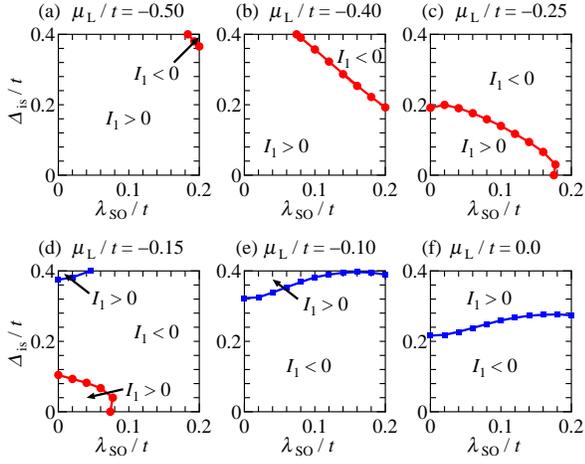}
\caption{Phase diagram reporting the $0$ and $\pi$-phases which are determined by evaluating 
the sign of the first harmonic term of the Josephson current $I_1$
for the NCS/NI/SSC of the (100) direction in the parameters space spanned by
the spin-orbit coupling ($\lambda_\mathrm{SO}/t$)  and the inversion symmetry breaking term ($\Delta_\mathrm{is}/t$). 
We consider the impact of the electron filling variation by determining the phase diagram for various values of the NCS chemical potential: 
(a)$\mu_\mathrm{L}/t=-0.50$, (b)$\mu_\mathrm{L}/t=-0.40$, (c)$\mu_\mathrm{L}/t=-0.25$, (d)$\mu_\mathrm{L}/t=-0.15$, 
(e)$\mu_\mathrm{L}/t=-0.10$, and (f)$\mu_\mathrm{L}/t=0.0$. 
The other parameters are set at $t_\mathrm{int}=1.0$ and $T=0.99T_\mathrm{cL}$. }
\label{fig5_v2}
\end{figure}%

\begin{figure*}[htbp]
\centering
\includegraphics[width=11cm]{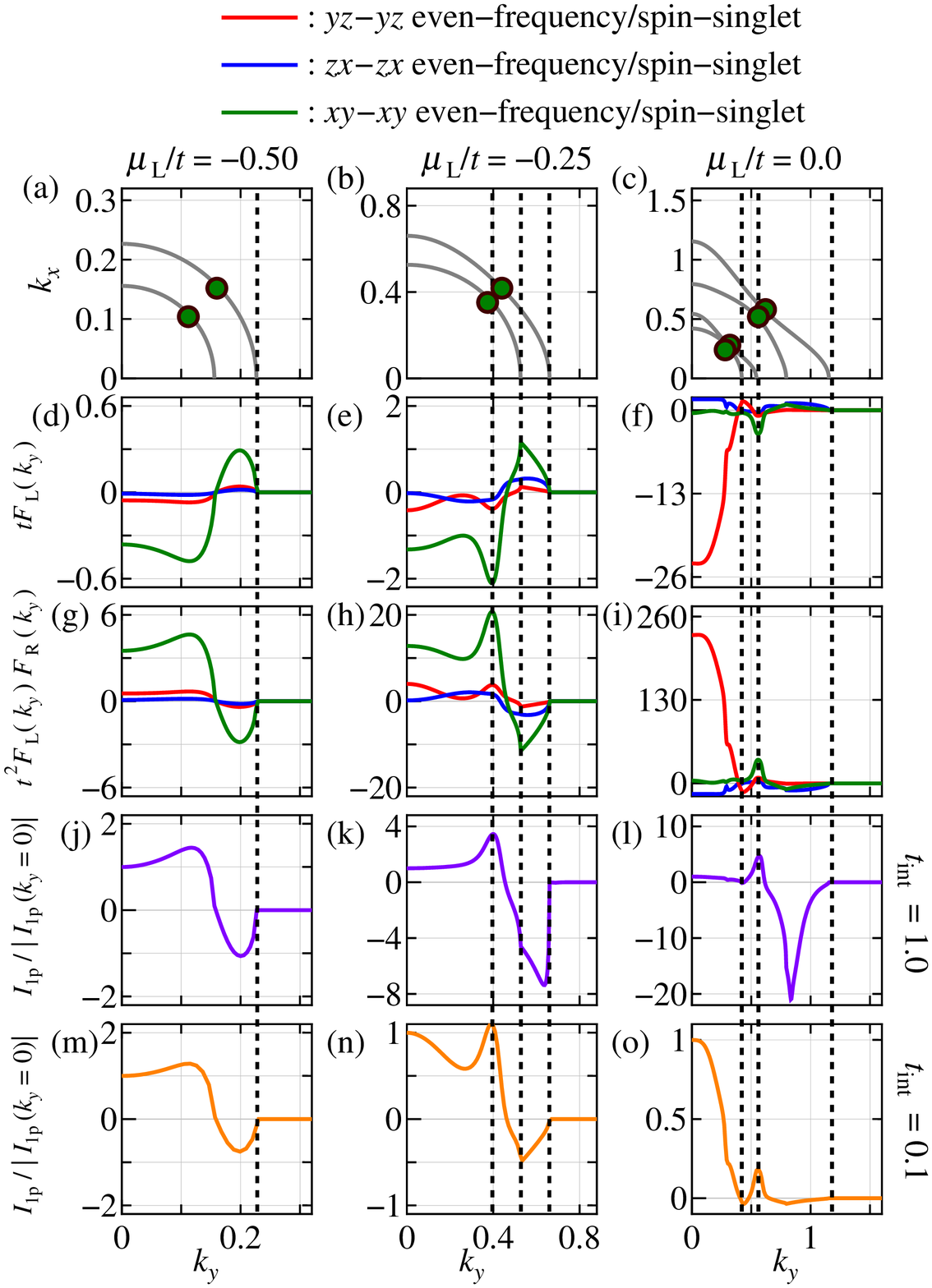}
\caption{Fermi surface in the NCS 
at (a)$\mu_\mathrm{L}/t=-0.50$, (b)$\mu_\mathrm{L}/t=-0.25$, and (c)$\mu_\mathrm{L}/t=0.0$. 
Pair amplitude of intraorbital spin-singlet component for each $t_{2g}$-orbital in the NCS
$F_\mathrm{L}(k_y)=F^{(\alpha,\alpha)}_{\uparrow\downarrow-\downarrow\uparrow}$ ($\alpha=yz,zx,xy$)
at (d)$\mu_\mathrm{L}/t=-0.50$, (e)$\mu_\mathrm{L}/t=-0.25$, and (f)$\mu_\mathrm{L}/t=0.0$. 
Product of $F_\mathrm{L}$ with $F_\mathrm{R}$ where $F_\mathrm{R}(k_y)=F_{\uparrow\downarrow-\downarrow\uparrow}$  
is the spin-singlet pairing amplitude in the single orbital $s$-wave SC
at (g)$\mu_\mathrm{L}/t=-0.50$, (h)$\mu_\mathrm{L}/t=-0.25$, and (i)$\mu_\mathrm{L}/t=0.0$, respectively. 
We note that $F_\mathrm{L}$ and $F_\mathrm{R}$ are calculated in the semi-infinite systems.
Momentum resolved first harmonic term $I_\mathrm{1p}(k_y)$ of the Josephson current normalized by $I_\mathrm{1p}(k_y=0)$
for high trasparency ($t_\mathrm{int}=1.0$) in (j)-(l) and low transparency ($t_\mathrm{int}=0.10$) in (m)-(o), respectively. 
In these panels, we select the chemical potential of the NCS 
as $\mu_\mathrm{L}/t=-0.50$ for (j)(m), $\mu_\mathrm{L}/t=-0.25$ for (k)(n), and $\mu_\mathrm{L}/t=0.0$ for (l)(o). 
Here, we obtain the first harmonic term $I_\mathrm{1p}(k_y)$ by the summation over the Matsubara frequency
at $i\varepsilon_n=-i\pi k_\mathrm{B}T$ and $i\pi k_\mathrm{B}T$. 
The other parameters are $\lambda_\mathrm{SO}/t=0.10$, $\Delta_\mathrm{is}/t=0.20$, and $T=0.10 T_\mathrm{cL}$. }
\label{fig6_v2}
\end{figure*}%

Next, we study the first harmonic ($I_1$) contribution to the Josephson current in the (100) direction
as a function of the spin-orbit coupling $\lambda_\mathrm{SO}$ and orbital Rashba interaction $\Delta_\mathrm{is}$ 
in the regime of high transparency 
since we have seen that only in that case one can observe a $0$-$\pi$ phase transition.
Apart from the role of the electron filling of the various bands, 
it is important to assess whether a variation of the electronic parameters associated with 
the strength of the spin-orbital entanglement 
and of the inversion asymmetry breaking can be employed to drive the 0- to $\pi$-phase transition.
The outcome is remarkable and unveils an intricate interplay between the band occupation 
(i.e.\ the orbital character of the Fermi surfaces) and the combination of $\lambda_\mathrm{SO}$ and $\Delta_\mathrm{is}$. 
In Fig.~\ref{fig5_v2} we present the resulting phase diagram constructed 
by evaluating the sign of the first harmonic term $I_1$ in the Josephson current in each point of the parameters space. 
We notice that there can be one or two boundaries that separate the 0 $(I_1>0)$ from the $\pi$-phase $(I_1<0)$ region 
in the parameters space $(\lambda_\mathrm{SO},\Delta_\mathrm{is})$. 
This implies that a reentrant type of 0-$\pi$ transition can be also obtained. 
For instance, 
by increasing the orbital Rashba coupling at $\mu_\mathrm{L}=-0.15$ for values of the $\lambda_\mathrm{SO}$ lower than about $0.10 t$, 
one can achieve a 0-$\pi$-0 changeover of the Josephson CPR. 
Another trend that can be deduced by inspection of the phase diagram is that the increase of the chemical potential $\mu_\mathrm{L}$ moves 
or generates 0-$\pi$ phase boundaries. 
The 0-$\pi$ boundary (red line in Fig.~\ref{fig5_v2}) shrinks towards the point $(\lambda_\mathrm{SO},\Delta_\mathrm{is})=(0,0)$ 
by increasing the chemical potential. On the other hand, 
at higher values of the electron filling, another boundary (blue line in Fig.~\ref{fig5_v2}) occurs 
at a lower threshold of the orbital Rashba coupling $\Delta_\mathrm{is}$. 
This phenomenon can be mainly ascribed to the $t_{2g}$-orbital components 
and the anisotropy of the spin-split Fermi surfaces with both nonzero $\lambda_\mathrm{SO}$ and $\Delta_\mathrm{is}$. 
It is particularly relevant to observe that in the low electron density regime, 
with only two Fermi surfaces and dominant $d_{xy}$ character, 
the $\pi$-phase can be achieved only for enough large $\lambda_\mathrm{SO}$ and $\Delta_\mathrm{is}$. 
Indeed, $\pi$-phase at $\mu_\mathrm{L}/t=-0.50$ appears at large $(\lambda_\mathrm{SO},\Delta_\mathrm{is})$ [Fig. \ref{fig5_v2}(a)]. 
The increase of the electron filling favors the interorbital mixing and the spin-orbital coupling can in turn activate the $\pi$-phase 
with smaller thresholds in the amplitude. 
When going through the Lifshitz transition~\cite{Lifshitz1960} from two to four Fermi surface electronic configuration, 
one observes an optimal regime for the $\pi$-phase that now covers almost the whole phase space 
in the explored $\lambda_\mathrm{SO}$ and $\Delta_\mathrm{is}$ amplitude. 
This outcome unveils the subtle role of the orbital degree of freedom in setting the $\pi$-state in the Josephson junction. 
Additionally, having found a 0-$\pi$ transition both in terms of a change in the electron filling 
and of the orbital Rashba coupling, we argue that this type of Josephson junction can manifest a dramatic response to an application of a gate voltage.
We note that the behavior in Fig.~\ref{fig5_v2} holds in the low temperatures 
since 0-$\pi$ transition does not occur by changing the temperature.

In order to get more insight into the origin of the sign change of the Josephson current in the (100) direction 
in terms of the variation of the chemical potential $\mu_\mathrm{L}$ in the regime of high transparency $t_\mathrm{int}=1.0$, 
we check the relation between the first harmonic term of the Josephson current $I_1$ in the (100) direction
and the induced intraorbital spin-singlet pair amplitude at the interface 
as a function of the conserved momentum $k_y$ (Fig.~\ref{fig6_v2}). 
The pair amplitude $\hat{F}_X$ is obtained by evaluating
\begin{align}
    \tilde{G}_X&=\frac{1}{i\varepsilon_{n}-\hat{H}^{X}_\mathrm{BdG}}\notag\\
    &=
    \begin{pmatrix}
        \hat{G}_X & \hat{F}_X \\
        \bar{F}_X & \bar{G}_X
    \end{pmatrix}.
\end{align}%
In the case of the three-orbital NCS (left-side SC),
the pair amplitude for the $(\alpha,\beta)$-orbitals $\hat{F}^{(\alpha,\beta)}_\mathrm{L}$ is described by
\begin{align}
    \hat{F}^{(\alpha,\beta)}_\mathrm{L}&=
    \begin{pmatrix}
        F^{(\alpha,\beta)}_{\uparrow\uparrow} & F^{(\alpha,\beta)}_{\uparrow\downarrow-\downarrow\uparrow}+F^{(\alpha,\beta)}_{\uparrow\downarrow+\downarrow\uparrow} \\
        -F^{(\alpha,\beta)}_{\uparrow\downarrow-\downarrow\uparrow}+F^{(\alpha,\beta)}_{\uparrow\downarrow+\downarrow\uparrow} & F^{(\alpha,\beta)}_{\downarrow\downarrow} \\
    \end{pmatrix},
\end{align}%
and the single orbital $s$-wave SC $\hat{F}_\mathrm{R}$, 
\begin{align}
\hat{F}_\mathrm{R}&=
\begin{pmatrix}
F_{\uparrow\uparrow} & F_{\uparrow\downarrow-\downarrow\uparrow}+F_{\uparrow\downarrow+\downarrow\uparrow} \\
-F_{\uparrow\downarrow-\downarrow\uparrow}+F_{\uparrow\downarrow+\downarrow\uparrow} & F_{\downarrow\downarrow} \\
\end{pmatrix}.
\end{align}%

In the Josephson junction upon examination, the spin-singlet pairing components 
in both left and right-side SCs can interfere
and contribute to the first harmonic term $I_{1}$ of the overall Josephson current. 
For this reason, it is useful to focus on the spin-singlet pair components 
and in particular to have a close inspection of their behavior at the junction's interface by computing the $k$-resolved amplitude. 
Here, $F^{(\alpha,\beta)}_{\uparrow\downarrow-\downarrow\uparrow}(k_y)$ refers to the NCS 
while $F_{\uparrow\downarrow-\downarrow\uparrow}(k_y)$ is for the spin-singlet amplitude 
in the single band $s$-wave SC.

As expected, the spin-singlet pair amplitude in the NCS is non-vanishing due to the combination of atomic spin-orbit coupling $\lambda_\mathrm{SO}$
and orbital Rashba interaction $\Delta_\mathrm{is}$. 
Since the intraorbital components are larger than the interorbital ones regarding the B$_1$ representation, 
the behavior of the intraorbital terms is more relevant for evaluating their role in setting out the Josephson current.  
The analysis has been conducted with the aim to identify the driving mechanisms or key physical quantities behind the formation of the $\pi$-state 
in the junction. 
As we have seen in the previous section, 
the intraorbital spin-singlet pair amplitude in the bulk has a sign change 
on the inner and outer Fermi surfaces with $d$-wave pattern. 
Then the CPRs which come from the outer and inner Fermi surfaces in the bulk NCS compete each other. 
This kind of cancellation has been proposed in iron-based $s_{\pm}$ SC/canonical SSC Josephson junction~\cite{Burmistrova2015PRB}. 
Moreover, a closer inspection of the amplitude distribution in the momentum space reveals a sublte anisotropy. 
Indeed, for the lowest electron filling ($\mu_\mathrm{L}/t=-0.50$) 
the strength of the spin-singlet pairing is larger along the $k_x$ or $k_y$ symmetry directions, 
while in the intermediate electron density, 
corresponding to $\mu_\mathrm{L}/t=-0.25$, the pair amplitude is more enhanced close to the diagonal directions. 
A similar behavior is also obtained for the $d_{xy}$ projected spin-singlet pairing at $\mu_\mathrm{L}/t=0.0$. 
For this electron filling, the $d_{zx}$ or $d_{yz}$-components, 
on the other hand, have a significant amplitude difference along the outer Fermi surfaces indicating that for those momenta the sign change 
cannot result into a complete cancellation when contributing to the Josephson processes.
Specific aspects that point to sign competition and anisotropy are also found 
for the $k_y$-projected intraorbital spin-singlet pair amplitude at the edge of the NCS 
close to the junction interface. 
We generally find that the intraorbital spin-singlet pair amplitude $F_\mathrm{L}$ 
tends to have a sign change for momenta $k_y$ [Figs.~\ref{fig6_v2}(d)-\ref{fig6_v2}(f)] 
that are in between those associated with the nodal points of the spin-triplet gap in the NCS [Figs.~\ref{fig6_v2}(a)-\ref{fig6_v2}(c)]. 
Moreover, $F_\mathrm{L}$ can have a high intensity for values of $k_y$ 
corresponding to the Fermi wave-vectors at $k_y=0$ or nearby the nodal points. 
Those momenta are characteristic of the nodal topological SCs 
and of the underlying Fermi surface.
In particular, it is useful to highlight the $k_y$ distribution 
of the intraorbital spin-singlet $F_\mathrm{L}$ amplitude. 
The outcome of the analysis indicates a strong orbital and electron filling dependence. 
The $d_{xy}$ component has comparable amplitude at small and large $k_y$ for $\mu_\mathrm{L}/t=-0.50$ and $\mu_\mathrm{L}/t=-0.25$, respectively, 
while for a higher electron filling (e.g.\ $\mu_\mathrm{L}/t=0.0$) the dominant spectral weight is distributed at large value of $k_y$ 
towards the position of the nodal points. 
On the other hand, the behavior of the $d_{zx}$ and $d_{yz}$ pairing amplitude is quite different from that of the $d_{xy}$. 
Indeed, the spectral distribution of the $d_{zx}$ indicates that the corresponding spin-singlet pairing amplitude is mostly contributing when
$k_y$ is close to the nodal points momenta. 
Hence, the behavior of the induced spin-singlet pair amplitude at the edge typically 
changes sign as a function of $k_y$ and its amplitude is strongly dependent on the orbital character and electron filling. 

With this know-how, we are ready to evaluate a possible link between the behavior of 
the $k$-resolved intraorbital spin-singlet pair amplitude 
with that of the first harmonic term of the Josephson current.
In particular, in the tunneling regime the product of the left and right intraorbital spin-singlet component 
$F_\mathrm{L}F_\mathrm{R}$ can be directly compared with the first harmonic Josephson term $I_1$. 
Indeed, for such configuration we have that $I_1 \sim F_\mathrm{L}F_\mathrm{R}$ 
as one can deduce by comparing the results in Figs.~\ref{fig6_v2}(g), \ref{fig6_v2}(h), and \ref{fig6_v2}(i) 
with those in Figs.~\ref{fig6_v2}(m), \ref{fig6_v2}(n), and \ref{fig6_v2}(o). 
The lack of a $\pi$-phase state emerges out of a subtle competition between the positive and negative Josephson channels 
when inspecting the $k$-resolved first harmonic term. 
Here, $I_{1p}(k_y)$ is obtained by the summation over the Matsubara frequency 
at $i\varepsilon_n=-i\pi k_\mathrm{B}T$ and $i\pi k_\mathrm{B}T$,
\begin{align*}
    I_1&\propto\sum_{k_y}I_\mathrm{1p}(k_y),\\
    I_\mathrm{1p}(k_y)&\sim I_\mathrm{1p}(k_y,-i\pi k_\mathrm{B}T)+I_\mathrm{1p}(k_y,i\pi k_\mathrm{B}T).
\end{align*}%
\begin{figure}[htbp]
\centering
\includegraphics[width=7.8cm]{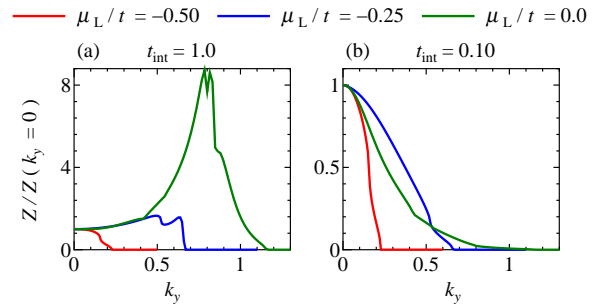}
\caption{Normalized charge conductance with normal metal configurations in the two sides of the junction 
    at (a)$t_\mathrm{int}=1.0$ and (b)$t_\mathrm{int}=0.10$. 
    Red, blue, and green lines denote the examined chemical potentials, i.e.\ $\mu_\mathrm{L}/t=-0.50$, 
    $\mu_\mathrm{L}/t=-0.25$, and $\mu_\mathrm{L}/t=0.0$, respectively. 
    We set the parameters as $\lambda_\mathrm{SO}/t=0.10$ and $\Delta_\mathrm{is}/t=0.20$.}
    \label{fig7_v2}
\end{figure}%

On the contrary, for high transparency, 
the behavior of $I_\mathrm{1p}(k_y)$ does not correlate with that of the intraorbital spin-singlet pairing amplitude product 
in Figs.~\ref{fig6_v2}(g), \ref{fig6_v2}(h), and \ref{fig6_v2}(i). 
Since the conductance at the high transparency is larger than that at the low transparency for large momentum
as shown in Fig.~\ref{fig7_v2}, 
$I_\mathrm{1p}(k_y)$ can be more affected by the contribution of multiple injection and reflection processes
for the various momenta. 
We find that the contributions of the large momentum regions to the Josephson current 
are those that allow to turn the sign from positive to negative 
when integrating the Josephson current over all of the momenta $k_y$.

\section{Temperature dependence of Josephson current}


In this section, we present the temperature dependence of the maximum Josephson current. 
The behavior of the maximum Josephson current for the temperature $T$ depends on 
the zero-energy surface ABSs at the interface's junction~\cite{Josephson1,Josephson2,Josephson3,KashiwayaTanaka2000RepProgPhys}.
In the present case, since the $s$-wave SC
does not have the surface ABSs at the edge,
we focus on the surface ABSs in the NCS.
For the (100) direction, the helical edge states appear in the case with two Fermi surfaces 
and the surface ABSs not connecting at the zero-energy appear in the case with four Fermi surfaces~\cite{fukaya18}.
In the (110) direction, zero-energy flat bands occur due to the topological properties of the nodal points in the NCS~\cite{fukaya18}.

\begin{figure}[htbp]
    \centering
    \includegraphics[width=7.8cm]{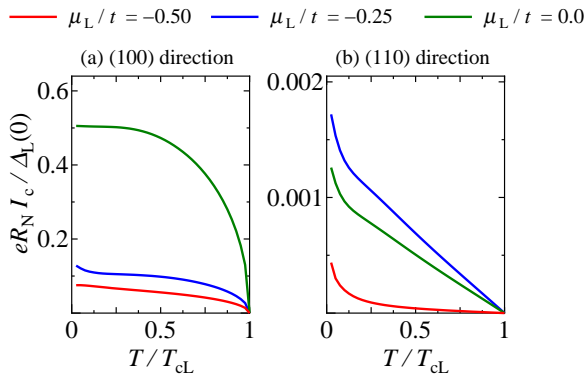}
    \caption{Temperature dependence of the maximum Josephson current 
    regarding the interorbital B$_1$ state in the NCS for the NCS/NI/SC junction. 
    (a)Temperature dependence of the maximum Josephson current in (100) junction 
    at $\mu_\mathrm{L}/t=-0.50$ (red line), $\mu_\mathrm{L}/t=-0.25$ (blue line), and $\mu_\mathrm{L}/t=0.0$ (green line). 
    (b)Temperature dependence of the maximum Josephson current in (110) junction 
    at $\mu_\mathrm{L}/t=-0.50$ (red line), $\mu_\mathrm{L}/t=-0.25$ (blue line), and $\mu_\mathrm{L}/t=0.0$ (green line). 
    The parameters are $\lambda_\mathrm{SO}/t=0.10$, $\Delta_\mathrm{is}/t=0.20$, and $t_\mathrm{int}=1.0$. }
    \label{fig8}
\end{figure}%
For these zero-energy surface ABSs, 
we can expect that the maximum Josephson current increases as the temperature is reduced~\cite{Josephson1,Josephson2,Josephson3,KashiwayaTanaka2000RepProgPhys}.
Fig.~\ref{fig8}(a) shows the temperature dependence of the maximum Josephson current 
at $\mu_\mathrm{L}/t=-0.50$ (red line), $\mu_\mathrm{L}/t=-0.25$ (blue line), and $\mu_\mathrm{L}/t=0.0$ (green line)
in the (100) direction. 
At $\mu_\mathrm{L}/t=-0.50$ and $\mu_\mathrm{L}/t=-0.25$, the Josephson current tends to increase,
however, at $\mu_\mathrm{L}/t=0.0$ its amplitude saturates at low temperature.
As we have shown in the previous section,
intraorbital even-frequency/spin-singlet/$d^{\pm}_{x^{2}-y^{2}}$-wave pair amplitude 
can be coupled to spin-singlet $s$-wave state thus directly affecting the Josephson current.
The emergent properties of the intraorbital even-frequency/spin-singlet/$d^{\pm}_{x^{2}-y^{2}}$-wave components [Fig.~\ref{fig5_v2}(f)] 
can also determine the behavior of the temperature dependence of the maximum Josephson current.
In the (100) direction, since the sign of the intraorbital even-frequency/spin-singlet/$d^{\pm}_{x^{2}-y^{2}}$-wave pair amplitude 
does not change at the interface,
the thermal behavior of the Josephson current is not influenced by the spin-singlet pair amplitude.
As a result, the Josephson current increases at low temperature 
due to the zero-energy surface ABSs at $\mu_\mathrm{L}/t=-0.50$ and $\mu_\mathrm{L}/t=-0.25$,
and is saturated by no zero-energy surface ABSs at $\mu_\mathrm{L}/t=0.0$ \cite{Asano2011PRB}.

Likewise, we determine the temperature dependence of maximum Josephson current in the (110) direction
as shown in Fig.~\ref{fig8}(b)
at $\mu_\mathrm{L}/t=-0.50$ (red line), $\mu_\mathrm{L}/t=-0.25$ (blue line), and $\mu_\mathrm{L}/t=0.0$ (green line).
At low temperature, the Josephson current shows a rapid upturn
owing to the zero-energy surface ABSs.
These zero-energy surface ABSs indicate that the sign of the intraorbital even-frequency/spin-singlet/$d^{\pm}_{x^{2}-y^{2}}$-wave pair amplitude 
changes for processes associated with the (110) direction.
Thus, the Josephson current in the (110) direction increases at very low temperature
due to the anisotropy of the intraorbital even-frequency/spin-singlet/$d^{\pm}_{x^{2}-y^{2}}$-wave pair amplitude of 
the interorbital B$_1$ pairing~\cite{Josephson1,Josephson2,Josephson3,KashiwayaTanaka2000RepProgPhys}.

\section{Conclusions and discussion}

We study a Josephson junction made of an NCS with local interorbital spin-triplet pairing interfaced 
with a conventional spin-singlet $s$-wave SC by considering different junction's orientation 
and exloring the various regimes of electron filling and spin-orbital coupling. 
We demonstrate that this type of superconducting pairing leads to a sign-changing intraorbital spin-singlet pair amplitude 
on different bands with $d$-wave symmetry. 
Such multi-band $d^{\pm}$-wave state is responsible of unexpected Josephson effects with 0-$\pi$ transitions 
displaying a high degree of electronic control. 
Remarkably, we find that the phase state of a NCS/NI/SSC Josephson junction 
can be switched between 0 and $\pi$ in multiple ways through a variation of electron filling, strength of the spin-orbital coupling, 
amplitude of the inversion asymmetry interaction, junction orientation and transparency. 
These results highlight an intrinsic orbital and electrical tunability of the Josephson response especially 
when considering the variation of the orbital Rashba coupling due to an applied electric field.

The presented results can find application in quantum materials where the electronic structure 
is marked by a strong interplay of spin and orbital degrees of freedom. 
This is commonly encountered in transition metal oxides and in particular at oxide interfaces or surfaces. 
A paradigmatic example is provided by the two-dimensional electron gas forming at the LAO-STO interface~\cite{Ohtomo2004Nature,Reyren2007Science}. 
There, the transport properties of a suitably designed Josephson junction reveal 
the presence of competing 0- and $\pi$-channels \cite{Tafuri2017PRBR}. 
We argue that the interorbital pairing here studied can account, at least qualitatively, 
for the observed anomalies and the Josephson phase frustration as a consequence of the nontrivial surface ABSs 
arising from both the spin-triplet and spin-singlet pairing components. 

Finally, we have proposed the spin-orbitronics functionalities to control the 0-$\pi$ transitions in Josephson devices. 
In particular, the remarkable tunability of the Josephson effect by means of electron filling, 
orbital Rashba interaction and the interface's transparency indicate several ways towards an electrical design of Josephson devices 
by directly gating the SC or by gating the interface.


\begin{acknowledgements}
    This work was supported by the JSPS Core-to-Core program ``Oxide Superspin'' international network, 
    and a JSPS KAKENHI (Grants No.\ JP15H05851, No.\ JP15H05853, No.\ JP15K21717, No.\ JP18H01176, No.\ JP18K03538, No.\ JP20H00131, and
    No.\ JP20H01857), 
    and the project Quantox of QuantERA-NET Cofund in Quantum Technologies, implemented within the EU-H2020 Programme. 
    Y.\ F.\ is supported by a JSPS research fellowship and JSPS KAKENHI (Grants No.\ 19J11865).
    M.\ C.\ and P.\ G.\ acknowledge support by the project ``Two-dimensional Oxides Platform for SPIN-orbitronics nanotechnology (TOPSPIN)''
    funded by the MIUR Progetti di Ricerca di Rilevante Interesse Nazionale (PRIN) Bando 2017 - Grant No.\ 20177SL7HC.    
\end{acknowledgements}

\bibliographystyle{apsrev4-1}
\bibliography{main}
\end{document}